\documentstyle[12pt%
,cite,epsfig%
]{article}
\def\Rf#1#2#3#4{{#1} {#2} (19#3) #4}

\def\PL{{Phys. Lett.}  B}
\def\PRp{Phys. Reports}
\def\PRC{{Phys. Rev.} C}

\def\PRL{Phys. Rev. Lett.}
\def\ZP{{Z. Phys.} C}
\def\YaF{Sov. J. Nucl. Phys.}
\def\JP{{J. Phys.} G}

\def\ea{\em et al.}
\def\ct{\cite}

\def\be{\begin{equation}}
\def\ee{\end{equation}}
\def\bfg{\begin{figure}}
\def\efg{\end{figure}}
\def\la{\label}
\def\bcr{\begin{center}}
\def\ecr{\end{center}}
\def\vs{\vspace}
\def\bi{\bibitem}
\def\fn{\footnote}

\def\rd{\rm d}
\def\e{\eta}

\def\ve{\varepsilon}
\def\vt{\vartheta}
\def\ef{\stackrel{\sim}{\e}}
\def\dep{\delta$${\ef}}
\def\ecf{\ef_{\rm 0}}
\def\al{\langle}
\def\ar{\rangle}
\def\De{\Delta\e}

\def\dn{\delta n}
\def\r{\rho}
\def\ddf{\r(\ef)}
\def\mn{\rm min}
\def\mx{\rm max}

\def\c2{\chi^2/{\rm DOF}}
\def\st{(\rm stat)}
\def\sy{(\rm syst)}

\begin{document}

\pagestyle{empty}
%%%%%%%%%%%%%%%%%%%%%%%%%%%%%%%%%%%%%%%%%%%%%%%%%%
%                                                %
%                  TITLE LIST                    %
%                                                %
%%%%%%%%%%%%%%%%%%%%%%%%%%%%%%%%%%%%%%%%%%%%%%%%%%
%
\begin{center}
{\Large \bf
On coherent particle production\\ 
\bigskip
in central 4.3 $A$ Gev/$c$ Mg-Mg collisions
}
\bigskip

\bigskip

{\large G.L. Gogiberidze$^{\rm a,}$\fn {On leave from Institute of
Physics, Tbilisi 380077, Georgia.}, L.K. Gelovani$^{\rm a,1,}$\fn {Now at
Indiana University Cyclotron Facility, Bloomington IN 47408, USA.},
E.K. Sarkisyan$^{\rm b}$} \\
\medskip

$^{\rm a}$~{\it Joint Institute for Nuclear Research, 
Dubna 141980, Moscow Region, Russia}

$^{\rm b}~${\it School of Physics \& Astronomy,
Tel Aviv University, Tel Aviv 69978, Israel}
\vs*{1.8cm}
%%%%%%%%%%%%%%%%%%%%%%%%%%%%%%%%%%%%%%%%%%%%%%%%%%%%%%%%%%%%%%
\end{center}
\medskip
\centerline{\large \bf Abstract}
\vs*{1.cm}

\noindent
Features of dense groups, or spikes, of negative pions produced in
Mg-Mg collisions at 4.3 GeV/$c$/nucleon are studied to search for a
coherent, {\v C}erenkov-like, mechanism of particle production process. 
We investigate the distributions of spike centers and, for the first time,  
the energy spectra of particles in spikes.
The spike-center distributions are obtained to exhibit the structure
due to the coherent gluon-jet emission dynamics. This structure is similar
to that observed recently for all-charged-particle spikes in hadronic and
nuclear interactions.
The energy distribution within spikes is found to have a significant
peak
over the inclusive background, while the inclusive spectrum shows
exponential decrease with two characteristic values of average kinetic
energy.
The value of the peak energy and its width are in a good agreement with
those expected for pions produced in a nuclear medium in the framework of
the {\v C}erenkov quantum approach.
The peak energy obtained is consistent with the value of the
cross-section maximum observed in coincidence experiments of
nucleon-nucleus interactions.

%%%%%%%%%%%%%%%%%%%%%%%%%%%%%%%%%%%%%%%%%%%%%%%%%%
%                                                %
%     END OF THE TITLE LIST                      %
%                                                %
%%%%%%%%%%%%%%%%%%%%%%%%%%%%%%%%%%%%%%%%%%%%%%%%%%

\newpage
\pagestyle{plain}
\setcounter{page}{1}

%%%%%%%%%%%%%%%%%%%%%%%%%%%%%%%%%%%%%%%%%%%%%%%%%%
%                                                %
%    BEGINNING OF TEXT                           %
%                                                %
%%%%%%%%%%%%%%%%%%%%%%%%%%%%%%%%%%%%%%%%%%%%%%%%%%

\section{Introduction}

A coherent component of particle production mechanism, {\v C}erenkov-like
radiation, in high energy particle collisions has been introduced a long
time ago. The idea of mesonic \ct{ivan} and scalar (pionic) \ct{glas}
radiation has recently been a subject of the systematic analysis of
production
of mesons in high-energy pion-nucleon scattering in nuclear medium in
terms of classical quantum mechanics \ct{ions}. Characteristic signatures
of the {\v C}erenkov mechanism, such as the differential cross-sections
and  angle-energy correlations of produced particles, have been
predicted.

Another approach of the {\v C}erenkov-like radiation in strong
interactions was suggested within the QCD based coherent gluon-jet
emission model \ct{revd}. In this model the pseudorapidity distributions 
of
centers of particle dense groups, called spikes,
are proposed to be investigated. The distributions of spike centers are
predicted to have two peaks due to destructive interference for quarks of
the same colour (pp collisions) or to be singly peaked due to constructive
interference for quarks of different colour (e.g. p$\bar {\rm p}$
interactions).  Recent observations in hadronic \ct{hh} and nuclear
\ct{plgog} interactions are found to be in agreement with these
predictions.

In this letter we search for dynamics of spike formation process 
using negative pions from collisions of relativistic nuclei.  The finite
size gluon-jet production mechanism is applied to obtain the coherent
dynamics, while further analysis deals  with the energy spectrum of
emitted pions in the framework of the nuclear pionic {\v C}erenkov-like
radiation (NPICR) approach. 

It is worth to mention that spikes have been extensively
investigated last years using stochastic picture of particle production
mechanism, namely intermittency phenomenon has been searched for
and obtained
in all types of collisions \ct{revi}. In our studies \ct{we}, we also
found the intermittency effect leading to multifractality and, then, to a 
conclusion of a
possible non-thermal phase transition during the cascading.  The latter
observation have been
confirmed in different reactions \ct{wec}.  

\section{Experimental details}

The results presented here are based on the experimental data obtained
after processing the pictures from the 2m Streamer Chamber SKM-200
\ct{skm} with a magnesium target placed inside. The chamber was installed
in a 0.8~T magnetic field and it was irradiated with a beam of
relativistic magnesium nuclei with momentum 4.3 GeV$/c$ per nucleon at the
Dubna JINR Synchrophasotron. A trigger selected central collisions has
been used in the experiment. The trigger started the Chamber if there were
no charged or neutral projectile fragments (momentum per nucleon required
to be greater than 3 GeV$/c$) emitted in a forward cone of 2.4$^{\circ}$.
A more detailed description of the set-up design and data reduction
procedure are given elsewhere \ct{skm,skm1,idjmp}.  Systematic errors
related to the trigger effects, low-energy pion and proton detection, the
admixture of electrons, secondary interactions in the target nucleus etc.
have been considered in detail earlier and the total contribution is
estimated do not exceed 3\% \ct{skm1,skm2}.

A total of 14218 Mg-Mg events were found to meet the
above centrality criterion. 
In the utilized sample only negative charged particles
(mainly $\pi^-$ mesons with a portion of some 1\% kaons) have been
studied. 
The average measurement error in momentum
$\al\ve_p/p\ar $ was about 1.5\% and that in the production angle
determination was $\al\ve_{\vt}\ar\simeq 0.1^{\circ}$.  The particles were
selected in the pseudorapidity ($\e=-\ln\tan\frac{1}{2}\vt$)  window of
$\Delta \eta =0.4 -2.4$ (in the laboratory frame) in which the angular
measurement accuracy does not exceed 0.01 in $\e$ units. The mean
multiplicity of the selected pions is $6.70\pm 0.02$.

The choice of $\pi^-$ mesons for the analysis presented is due to the fact
that they dominate among the produced particles and are well identified.
On the other hand, the production of pions is predominant process at the
energies of the Dubna Synchrophasotron and, therefore, they carry
important information about the dynamics of collisions.  The inclusive
characteristics of $\pi^-$'s in the reaction under investigation have been
studied earlier in Ref. \ct{idjmp}. Increase of statistics by more than
twice
allows analysing of more detail features of particle production
process.

To overcome an influence of the shape of the pseudorapidity distribution
on the results we  use the ``cumulative variable'',

\be
\ef(\e)\; =
\int_{\e _{\mn}}^{\e}
\r(\e ')\, {\rd} \e ' \,
\Bigg/
\int_{\e _{\mn}}^{\e _{\mx}} \r(\e ')\, {\rd} \e '\; ,
\la{nv}
\ee

\noindent
with the uniform spectrum $\ddf$ within the interval [0,1], as advocated
in Ref. \ct{fl}. This transformation makes possible to
compare results from different experiments.

The spikes are extracted in each event from the ordered pseudorapidities
which are scanned with a fixed pseudorapidity interval (bin) of size
$\dep$. The spikes with definite number of particles $\dn$, hit in
the bin, are determined and then the distributions of centers of spikes,
averaged over all events, are obtained. The center of spike is defined by   
$\ecf\,=(1/\dn)\sum_{j=1}^{\dn}\ef_j$.

To reveal dynamical correlations, the $\ecf$-distribution is
compared with analogous distributions obtained from the simulated
pseudorapidity single-particle spectrum $\ddf$ without any input
information about particle correlations.  The simulation procedure was as
follows.  According to the multiplicity distribution of the data sample,
the number of particles were randomly generated. Then, the
pseudorapidities
were distributed in accordance with the experimental $\ef$-spectrum and
corresponding to the generated multiplicity. The total number of the
simulated events was about 1.5 million, much more than 100 times the
experimental events.  Evidently, the obtained sample represents 
results from independent particle emission process.

\section{The results}

\subsection{Spike-center distributions}

\bfg[t]
\bcr
\epsfig{figure=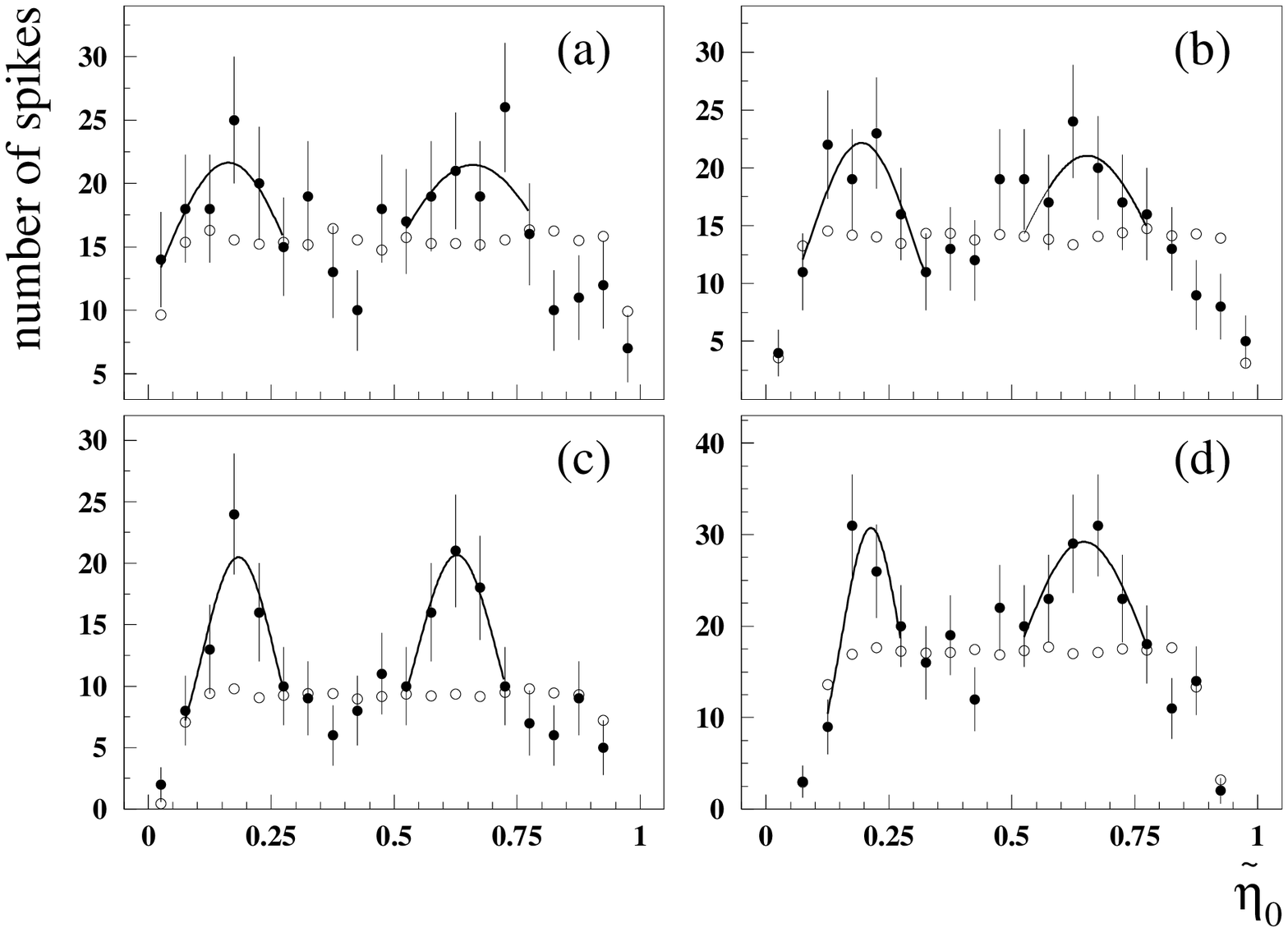,height=3.71in,
}
\ecr
\caption{\small
Experimental ($\bullet$) and simulated ($\circ$)
spike-center distributions for  different $\dep$-bins and
$\dn$-multiplicities:
{\bf (a)} $\dep\, = 0.05,\; \dn=4$, {\bf (b)} $\dep\, = 0.1,\; \dn=5$,
{\bf (c)} $\dep\, = 0.15,\; \dn=6$, {\bf (d)} $\dep\, =  0.25,\; \dn=7$.
The curves represent Gaussian fits.
}
\efg

Fig. 1 shows the pseudorapidity spike-center $\ecf$-distributions for four
different size $\dep$-bins and for spikes of various multiplicities $\dn$.
A two-peak
structure of the measured distributions (solid circles) with the peaks in
the neighbourhood of the same $\ecf$ is seen independent of the width and
multiplicity of spike. The shape of the distributions is in agreement with
the structure predicted by the coherent gluon-jet emission model \ct{revd}
and is similar to that observed earlier in hadronic interactions \ct{hh}
and by us in C-Cu collisions \ct{plgog}.

In order to estimate the positions of the peaks and the distance between
them, we fit these two bumps with Gaussians and average over
different spikes. The peaks are found to be placed at $\ecf\, \approx
0.19$
and 0.63, or at $\e_0= 0.89\pm0.03\st\pm0.08\sy$ and
$1.63\pm0.05\st\pm0.10\sy$.  They are separated by the $d_0$ interval,

\be
d_0=0.75\pm0.06\st\pm0.13\sy
\la{d}
\ee
in $\e$ units.
This value is close to those from the above mentioned nuclear and
hadronic interactions.

\bfg[t]
\bcr
\epsfig{figure= 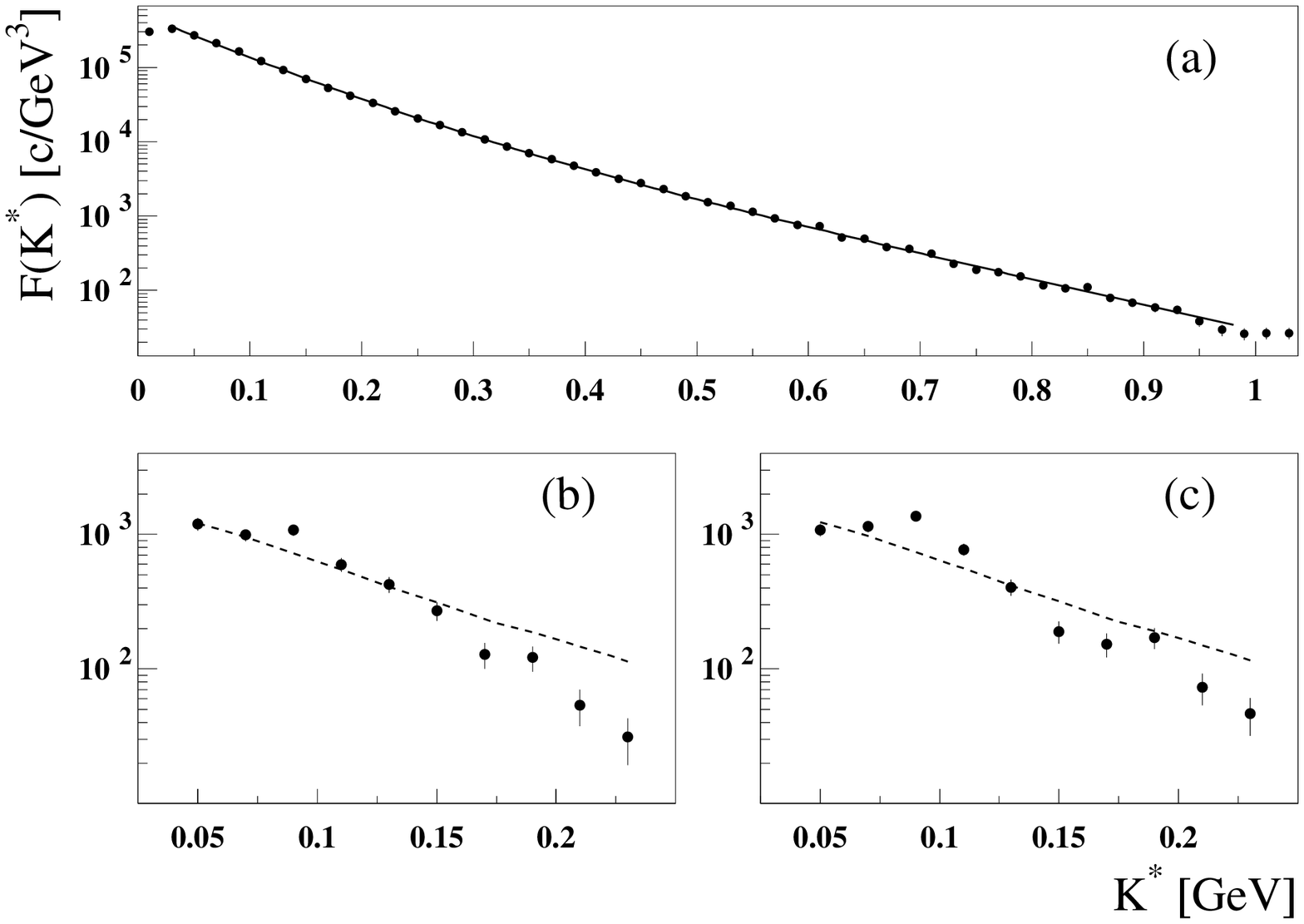,height=3.71in,
}
\ecr
\caption{\small
        Inclusive kinetic energy distribution {\bf (a)} 
        and    
        analogous distributions for spikes of
        {\bf (b)} $\dep\, = 0.1,\; \dn=6$ 
        and  
        {\bf (c)} $\dep\, = 0.15,\;\dn=7$. 
The solid line represents the exponential fit with Eq. (\ref{teme}), 
the dashed lines show the inclusive background. 
}
\efg

The dynamical origin of the structure obtained comes from a comparison of
the experimental $\ecf$-distributions with those based on the
above described simulated
events (open circles in Fig. 1). No peaks are seen in the simulated
distributions levelling off at the background, far away from the
measured
peaks. This points to the dynamical effect in the formation of spikes due
to 
the coherent gluon radiation picture.

In order to assess the reliability of the results obtained, we studied the
influence of the $\De$-range used and the experimental error
$\al\ve_{\vt}\ar$ in the measurements of the polar angle $\vt$. Varying
the $\De$ range and the error $\al\ve_{\vt}\ar$ we found the structure of
the distributions unchanged and the positions of the two peaks and the
distance $d_0$ to be within the above shown errors, in support of the
conclusions made.

\subsection{In-spike energy spectra}

The obtained strong signal of the coherent emission dynamics allows
further search for its manifestations in the energy distributions as it is
predicted in the NPICR approach \ct{ions}. In this model, the energy
spectra of pions emitted through the coherent {\v C}erenkov-like mechanism
when a few GeV proton passes the nuclear medium is predicted to have a
peak. This peak is expected to appear at 260 MeV when absorption effect is
neglected and at 244 MeV otherwise.

In Fig. 2 we compare the c.m.s. inclusive kinetic energy distribution,
$F(K^*)$ $=(1/E^*p^*) \,{\rd}N/{\rd}K^*$, with analogous spectra 
calculated for pions from 
spikes. Here, $E^*$ and $p^*$ denote, respectively, the particle energy
and momentum in the c.m.s. frame.

Using the temperature description, utilized in the last time to
characterize a system of excited hadrons \ct{idjmp, broc, nagu, backo}, we
parametrize the inclusive spectrum (Fig. 2a) by a sum of two exponents,

\be
F(K^*) = A_1 \exp(-K^*/T_1)+A_2 \exp(-K^*/T_2)\, ,
\la{teme}
\ee

\noindent
where the temperatures $T_{1}<T_{2}$ characterize \ct{broc} the two
possible
mechanisms of pion production, via $\Delta$-resonance decay and directly,
and are related to the pion average kinetic energies. The range of the
parametrization
shown is limited from below and from above due to detector effects and
corresponding requirements on the momenta of pions. The fit gives $T_1= 65
\pm 1$ MeV
and $T_2= 127\pm 1$ MeV. These values are, in general, consistent with
those obtained from the earlier analysis of the reaction under study
\ct{idjmp} and from other experiments \ct{broc,nagu}.  Some difference in
the values could be explained if one takes into account the difference in
the sizes of the (pseudo)rapidity regions used \ct{broc}.

The shape of $F(K^*)$ distribution changes when the analysis is extended
to spikes, Figs. 2b and c.  The energy spectra of particles belonged to
spike differ significantly from the exponential law (\ref{teme}) and have
now a peaked shape. To extract the NPICR-signal we compare the
in-spike energy spectra with renormalized inclusive distribution, or
inclusive background, depicted with the dashed lines.  The first peak is
seen to be over the background with the statistical significance of 2.7
and 4.1 standard deviations in Figs. 2b and 2c, respectively. This peak is
located at the kinetic energy $K^* \approx 100$ MeV, or the total energy
$E^* \approx 240$ MeV, in accordance with the NPICR prediction.

To estimate the position of the peak and the bump width and to make the
results more comparable with the NPICR expectations, the
$E^*$-distributions of particles in spikes of various size $\dep$-bins
and different $\dn$-multiplicities have been studied.  Fig. 3. represents
some of
these distributions. The following specific peculiarities are found.
 
All these $E^*$-distributions possess a non-exponential behaviour with a
pronounced maximum in the vicinity of the value $E_{\rm m}^*=240$ MeV
regardless
the bin size and the multiplicity of spikes. Higher multiplicity of spike
is (at fixed $\dep$ size), more peaks appear.
The multi-peak structure is observed for bins with the multiplicities
$\dn
>3$, while at $\dn \leq 3$ only one peak occurs (not shown).

\bfg[t]
\bcr
\epsfig{figure=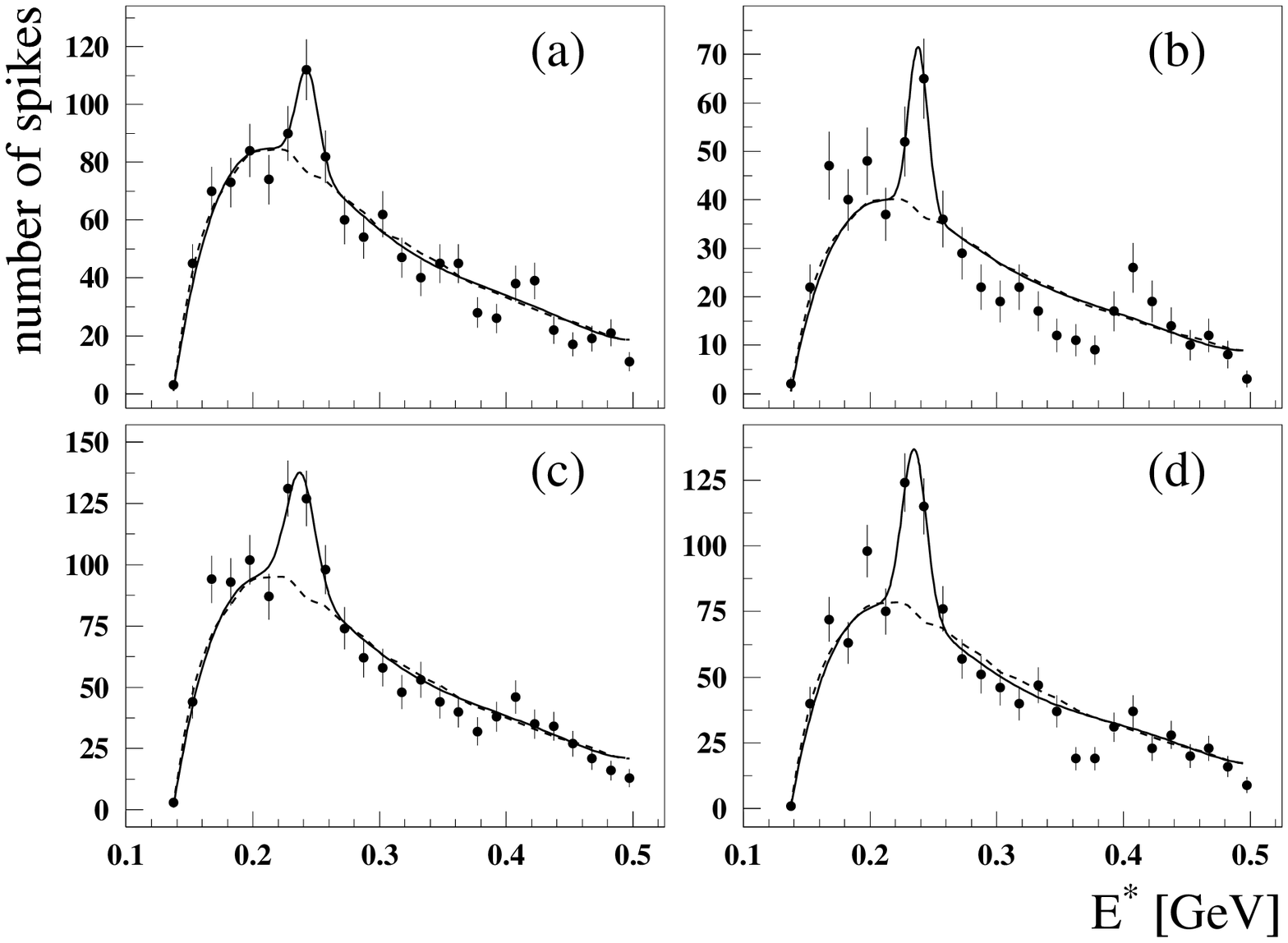,height=3.71in,
}
\ecr
\caption{\small
Total energy spectra for spikes of different $\dep$-bins and
multiplicities $\dn$:
{\bf (a)} $\dep\, = 0.05,\; \dn=4$, {\bf (b)} $\dep\, = 0.08,\; \dn=5$,
{\bf (c)} $\dep\, = 0.1,\; \dn=5$, {\bf (d)} $\dep\, =  0.15,\; \dn=6$.
The solid lines represent the fit (see text), the dashed lines show the
inclusive background.  
}
\efg

To reveal the dynamical signal we compare the in-spike energy
distributions
with the inclusive background (dashed-line). Alike to the above
kinetic energy distributions, the value of $E^*$ about 240 GeV is obtained
to be
most prominently and statistical-significantly peaked over the background.
To estimate the background and to parametrize the signal, we use a
fifth-order polynomial for the background and the Gaussian for the peak.
The solid curve shows the result of this fit.
After averaging over various spikes, the position of the peak and its
width are found to have the values,

\be
E^*_{\rm m}=238 \pm 3{\st}\pm 8{\sy} \; {\rm MeV}\, ,\;\;\;\; 
\Gamma_{\rm m}=10 \pm 3{\st}\pm 5{\sy} \; {\rm MeV} \, ,
\la{eg}
\ee

\noindent
respectively.

The location
of the obtained centre of the Gaussian is within the interval of the
energies
of pions expected from the {\v C}erenkov-like mechanism for
incident protons of a
few GeV, $224\leq E_{\rm m}\leq244$ MeV \ct{ions}. The value of
$E^*_{\rm m}$
(\ref{eg}) is similar to the position of the peak observed for
$\pi^+$ invariant mass distribution in analysis of coincidence
measurements of (p,n) reaction on carbon target at 1.5 GeV/c in the
$\Delta$-resonance excitation region \ct{chib}, the effect connected with
the NPICR mechanism \ct{ions}.
Also, the width $\Gamma_{\rm m}$ confirms an observation of the {\v
C}erenkov
radiation signal expected to be  $\Gamma \leq 25$ MeV.

\section{Conclusions}

Recapitulating, in order to search for a coherent, {\v C}erenkov-like
emission mechanism of particle production, a study of spikes in
relativistic nuclear collisions is carried out with negative pions from
central Mg-Mg collisions at a momentum of 4.3 GeV/$c$ per incident
nucleon. 
The spike-center distributions and, for the first time, the energy spectra
of particles within spike are investigated for various
narrow pseudorapidity bins and different spike multiplicities.

The spike-center distributions are found to possess a double-peak shape
that is in agreement with the structure expected from the coherent gluon
radiation model. 
The obtained distance between the peaks as well as  the shape of the
distributions are similar to those observed recently in analogous studies
of charged particle spikes in hadronic and relativistic nuclear interactions.
The dynamical effect in the spike-center distributions is revealed by
comparison with independent particle-emission model, where no peaks
are seen.

The coherent character of particle-production mechanism is confirmed by
studying energy distributions.  The inclusive energy spectra show
monotonic exponential decrease with two specific temperatures, while the
in-spike energy distributions are obtained to exhibit a peak at the
position
and of the width both consistent with the values expected from
the theoretical
calculations based on the
hypothesis of nuclear pionic {\v C}erenkov radiation. The value  of the
peak energy is close to the recently observed maximum in the
differential
cross-sections studied in the coincidence experiments of a few GeV
nucleon-nucleus interactions.

The results of the presented analysis signalize the coherent emission 
to be a complementary mechanism to the stochastic scenario of 
hadroproduction. Furthermore, the similarity of the 
spike-center distributions obtained for like-charged particles in the
presented paper with those for all-charged-particle spikes in the earlier
studies
indicates important contributions of the coherent mechanism
to formation of Bose-Einstein correlations \ct{bec}. It is worth to
mention that, in comparison to stochastic (intermittency) dynamics, 
the origin of which remains still unclear \ct{revi}, the coherent emission
has definite underlying dynamics.    
All this gives evidence for the necessity of further efforts
in studying existing experimental data.

\vs*{0.7cm}

\noindent{\Large \bf Acknowledgements}\\

We are grateful to the members of the GIBS (SKM-200) Collaboration for
providing us with the film data. 
The helpful discussions with V.A. Nikitin and his assistance are highly
acknowledged.


\small
\begin{thebibliography}{99}
\bibitem{ivan} W. Wada, Phys. Rev. 75 (1949) 981;
  D. Ivanenko and V. Gurgenidze, Dokl. Akad. Nauk SSSR 67 (1949) 997;
  D.I. Blokhintsev and V.L Indenbohm, ZhETF 20 (1950) 1123.
\bibitem{glas} W. Czy\.z and S.L. Glashow, Nucl. Phys. 20 (1960) 309; 
               G. Yekutieli, Nuovo Cim. 13 (1959) 446, 1306(E);
               P. Smr\v z, Nucl. Phys. 35 (1962) 165.
\bibitem{ions} D.B. Ion and W. Stoker, 
\Rf{\PRC}{48}{93}{1172}, \Rf{\PRC}{52}{95}{3332}, and refs. therein.
\bi{revd}
   I.M. Dremin, JETP Lett. 30 (1979) 140; Sov. J. Part. Nucl. 18 (1987) 31.
\bi{hh} I.M. Dremin {\ea}, \Rf{\YaF}{52}{90}{536}; 
        N.M. Agababyan {\ea}, EHS/NA22 Collab., \Rf{\PL}{389}{96}{397};
        S.-S. Wang, R. Liu, Z.-M. Wang, \Rf{\PL}{427}{98}{385}.
\bibitem{plgog} G.L. Gogiberidze, L.K. Gelovani, E.K. Sarkisyan,
                     \Rf{\PL}{430}{98}{368}.
\bi{revi} E.A. De Wolf, I.M. Dremin, W. Kittel, \Rf{\PRp}{270}{96}{1}.
\bi{we} E.K. Sarkisyan {\ea}, \Rf{\PL}{347}{95}{439}; 
         G.L. Gogiberidze {\ea}, Proc. 8th Int. Workshop on Multiparticle
         Production, 
         {\it Correlations and Fluctuations '98: From QCD to Particle
         Interferometry} (M\'atrah\'aza, 1998), 
         T.~Cs\"org\H o {\ea} (Eds.), World Scientific, 1999, p. 498.
\bi{wec}  D. Ghosh {\ea}, \Rf{\ZP}{71}{96}{243}; 
          S.-S. Wang, R. Liu, Z.-M. Wang, \Rf{\PL}{438}{98}{353}. 
\bi{skm} A. Abdurakhimov {\ea}, Instrum. Exp. Tech. 21 (1979) 1210.
\bi{skm1} M. Anikina {\ea}, \Rf{\PRC}{33}{86}{895}.
\bi{idjmp} L. Chkhaidze {\ea},  \Rf{\JP}{22}{96}{641}.
\bi{skm2}
      SKM-200 Collab., M. Anikina {\ea}, JINR report E1-84-785 (1984);
      JINR  Rapid Commun. 1[34] (1989) 12.
\bi{fl} A. Bia{\l}as and M. Gazdzicki, \Rf{\PL}{252}{90}{483};\\
                  W. Ochs, \Rf{\ZP}{50}{91}{339}.
\bibitem{broc} R. Brockmann {\ea},  \Rf{\PRL}{53}{84}{2012}.
\bi{nagu} S. Nagamia and M. Gyulassy, Adv. Nucl. Phys. 13 (1984) 201.
\bi{backo} S. Backovi\'c {\ea}, JINR Rapid Comm. 2[53] (1992) 58.
\bi{chib} J. Chiba  {\ea}, \Rf{\PRL}{67}{91}{1982}.
\bi{bec} R.M. Weiner, Phys. Reports (to appear), {\tt hep-ph/9904389}.
\end{thebibliography}
\end{document}